\documentclass[reprint,amsmath,amssymb,aps,prl]{revtex4-1}
\usepackage{graphicx}
\usepackage{dcolumn}
\usepackage{tensor}
\usepackage{amssymb}
\usepackage{amsmath}
\usepackage{bm}
\usepackage{mathtools}
\usepackage{tikz}

\newcommand\PlaceText[3]{%
\begin{tikzpicture}[remember picture,overlay]
\node[outer sep=0pt,inner sep=0pt,anchor=south west] 
  at ([xshift=#1,yshift=-#2]current page.north west) {#3};
\end{tikzpicture}%
}
\begin{document}
\preprint{APS/123-QED}
\title{Optical forces from near-field directionalities in planar structures}
\author{Jack J. \surname{Kingsley-Smith}}
\thanks{Corresponding author: jack.kingsley-smith@kcl.ac.uk}
\author{Michela F. \surname{Picardi}}
\author{Lei \surname{Wei}}
\author{Anatoly V. \surname{Zayats}}
\author{Francisco J. \surname{Rodr\'iguez-Fortu\~no}}
\affiliation{Department of Physics, King’s College London, Strand, London WC2R 2LS, United Kingdom}
\date{\today}

\begin{abstract}
Matter manipulation with optical forces has become commonplace in a wide range of research fields and is epitomized by the optical trap. Calculations of optical forces on small illuminated particles typically neglect multiple scattering on nearby structures. However, this scattering can result in large recoil forces, particularly when the scattering includes directional near-field excitations. Near-field recoil forces have been studied in the case of electric, magnetic and circularly polarized dipoles, but they exist for any type of directional near-field excitation. We use the force angular spectrum as a concise and intuitive analytical expression for the force on any dipole near planar surfaces, which allows us to clearly distinguish the effect due to the dipole, and due to the surface. We relate this directly to the coupling efficiency of surface or guided modes via Fermi's golden rule. To exemplify this, a near-field force transverse to the illumination is computationally calculated for a Huygens dipole near a metallic waveguide. We believe this formalism will prove insightful for various nanomanipulation systems within areas such as nanofluidics, sensing, biotechnology and nano-assembly of nanostructures.
\end{abstract}
\maketitle

\section{\label{sec:Intro}Introduction}

The manipulation of small objects using purely optical forces began with the optical trap \cite{Ashkin1970,Ashkin1986,Ashkin2000} and has since cemented itself as an important technique in a wide range of fields \cite{Grier2003,Neuman2004}. 
The trapping force in modern optical tweezers originates from the gradients of the electromagnetic field intensity around the trapped object and, typically, strong gradient forces are generated through the use of highly-focused beams and translating the focus to move the subject.

Tractor beams \cite{Marston2006,Dogariu2013,Gao2017,Sukhov2017} are a more recent optical manipulation concept which pulls the subject particle towards the illumination source with no equilibrium position. The concept has been proposed in a variety of configurations \cite{Lee2010,Sukhov2011,Brzobohaty2013,Ruffner2012,Ding2014}, including inducing a far-field directionality in the subject's multipolar forward scattering, using a Bessel beam or plane waves \cite{Chen2011,Novitsky2011}. This configuration produces an attractive, non-conservative force by increasing the ratio between the forward scattering and the backward scattering of the subject, constructing a directionality and a corresponding recoil force. 

The upscaling of optical tweezers and tractor beams to multiple objects is plagued by the complexity of aligning and controlling many laser beams. A recent addition to these nanomanipulation techniques, that has no such limitation, has been recently proposed in the form of surface recoil forces \cite{Rodriguez-Fortuno2015,Scheel2015,Sukhov2015}. These forces are similar to the recoil forces in tractor beams in that they are a result of directional scattering and momentum conservation. The difference is that they are applied to near-field excited guided waves \cite{Rodriguez-Fortuno2013,OConnor2014,Picardi2017a,Sinev2017,Wei2018,Bliokh2015,VanMechelen2016,Petersen2014,Neugebauer2014,leFeber2015,Shalin2014,Zhang2018,Picardi2018}. 
Recent works on these surface recoil forces include chiral sorting \cite{Wang2014,Hayat2015}, surface mode optical pulling \cite{Petrov2016} and lateral Casimir forces \cite{Manjavacas2017,Silveirinha2018}. 

This type of optical manipulation does not require incident fields with an electromagnetic gradient and instead utilizes the light-matter interactions that occur near the surface. Plane waves can, therefore, be used as a much simpler optical illumination which operates over a large region simultaneously and naturally leads towards scaled up systems. Since the recoil force is often polarization dependent, ultrafast polarization switching technologies \cite{Nicholls2017, Yang2017} can be implemented in the illumination beam path, allowing rapid control over the dynamics of the objects. The geometry of the system makes these forces highly applicable to lab-on-a-chip designs. Optical forces of this type could form part of the suggested future nanofactories and `bottom-up' fabrication techniques \cite{Raziman2015}. 

In this paper, we derive a general formalism for the force on any electromagnetic dipole within wavelength distance of a planar medium, which we call the angular force spectrum. Previous works on near-surface forces are limited to very specific scenarios (specific dipole moments, specific surfaces, particle chirality, etc) but are clearly all physically related. The angular force spectrum description is a robust and physically intuitive framework which is valid for any near-field directionalities in any type of planar structure, including plasmonic surface modes and dielectric planar waveguides. We also show that the excitation strength of the surface modes elegantly appears in the analytical expression for the force angular spectrum, through Fermi's golden rule. We use the Huygens dipole, recently predicted to exhibit near field directionality \cite{Kerker1983,Picardi2017a,Chaumet2009,Gomez-Medina2012,Aunon2014}, as a perfect example for which to computationally compare the magnitudes of near-field and far-field recoil forces. Finally, we compare these results with an optimized version of the well-known circular electric dipole \cite{Rodriguez-Fortuno2015,Rodriguez-Fortuno2013,Picardi2017a} to show that this near-field dominance of the force close to the surface is not specific to the Huygens but is instead an example of a general principle of surface recoil forces.

\section{\label{sec:level2}Near-Field recoil Optical forces}

The underlying principle of a recoil force is the conservation of linear momentum. If the scattering of an illuminated particle is directional, there is an imbalance in emitted radiation. Since the scattered light is carrying linear momentum away from the particle, the particle will experience a recoil force to conserve momentum. The harnessing of directional light as a means of producing recoil forces is a broadband effect and can be achieved in numerous ways. A preferential scattering in a direction  perpendicular to the illumination axis will produce a perpendicular force, therefore providing a theoretical mechanism for full control of the subject in 3D space.

The Maxwell stress tensor (MST) is a standard method of calculating optical forces and determines the rate of change of mechanical momentum within an arbitrary closed volume via a surface integral \cite{Novotny2006}. 
\begin{equation}
\langle\mathbf{F}\rangle = \int_{\mathcal{S}}^{} \langle\overset\leftrightarrow{\mathbf{T}}\rangle \cdot \textbf{\^n} \, d\mathcal{S}
\label{equ:MSTsurf}
\end{equation}

\noindent where $\mathbf{F}$ is the force acting on a body and $\textbf{\^n}$ is the normal vector perpendicular to and out of any arbitrary closed surface $\mathcal{S}$ enclosing the body. The MST  $\overset\leftrightarrow{\mathbf{T}}$ is defined as \cite{Jackson1999,Novotny2006,Chaumet2009}
\begin{equation}
\langle\overset\leftrightarrow{\mathbf{T}}\rangle = \frac{1}{2} \mathbb{R} \bigg\{\varepsilon \mathbf{E} \otimes \mathbf{E}^* + \mu \mathbf{H} \otimes \mathbf{H}^* - \frac{1}{2} \big(\varepsilon |\mathbf{E}|^2 + \mu |\mathbf{H}|^2\big)\overset\leftrightarrow{\mathbf{I}}\bigg\}
\label{equ:MST}
\end{equation}

\noindent where $\mathbf{E}$ and $\mathbf{H}$ are the total electric and magnetic fields, $\otimes$ denotes the outer product of two vectors, asterisks represent complex conjugations, $\overset\leftrightarrow{\mathbf{I}}$ is the identity matrix and $\varepsilon$ and $\mu$ are the permittivity and permeability of the medium, respectively. Throughout this paper, a time harmonic field dependence $\mathbf{E}(\mathbf{r},t) = \mathbb{R} \{ \mathbf{E}(\mathbf{r}) e^{-i\omega t} \} $ is assumed, where $\omega$ is the angular frequency of the field. In practice, the numerical surface integration can become computationally expensive due to the number of points necessary to acquire an accurate result. 

The system of interest to this paper is that of a generic magnetodielectric Rayleigh particle close to a planar surface. Under illumination, the particle can generate a electric dipole moment $\mathbf{p}$ and a magnetic dipole moment $\mathbf{m}$, which in the simplest case is given by:
\begin{align}\label{equ:polarizabilties}
\mathbf{p} &= \alpha_e \widetilde{\mathbf{E}} & \mathbf{m} &= \alpha_m \widetilde{\mathbf{H}}
\end{align}
\noindent where $\widetilde{\mathbf{E}}$ and $\widetilde{\mathbf{H}}$ are the total fields minus the dipole's self-fields. The self-fields are in some contexts called scattered fields, while the $\widetilde{\mathbf{E}}$ and $\widetilde{\mathbf{H}}$ fields are called incident or background fields. Their sum produces the total fields $\mathbf{E}$ and $\mathbf{H}$. The background fields are calculated at the location of the dipole moment (typically in the centre on the particle). $\alpha_e$ and $\alpha_m$ are the complex electric and magnetic polarizabilities, respectively. In this paper, we shall omit higher order multipoles for simplicity although the same recoil force principle applies. 

The total optical force acting on the particle is the sum of the radiative reaction force of the dipole fields and the Lorentz force that the background fields $\widetilde{\mathbf{E}}$ and $\widetilde{\mathbf{H}}$ exert on the dipole moments $\mathbf{p}$ and $\mathbf{m}$. However, this calculation is in general very complicated, because both the dipole moments and the background fields depend mutually on one another (via multiple scattering) giving rise to feedback mechanisms, resonances, etc... Fortunately, the problem can be split into two simpler logical steps: The first step is to find the dipole moments which satisfy Eq. (\ref{equ:polarizabilties}), including illumination and multiple reflections, and possibly including further complications such as gyrotropic or anisotropic particles which require generalized versions of Eq. (\ref{equ:polarizabilties}). This first step requires solving simultaneous equations self-consistently and is not the focus of this work. Once $\mathbf{p}$ and $\mathbf{m}$ have been calculated in this way, the second step of the problem is to find the force acting on the dipole moments. Fortunately, for this second step, we can work with the assumption that $\mathbf{p}$ and $\mathbf{m}$ are known values and study the force which the background fields $\widetilde{\mathbf{E}}$ and $\widetilde{\mathbf{H}}$ exert on them. This second part of the problem is also extremely interesting and is the focus of this paper, as the force becomes non-trivial in the presence of a nearby surface. Therefore we assume that the first step has been achieved and we work with $\mathbf{p}$ and $\mathbf{m}$ as if they were independent variables assumed to correspond self-consistently with Eq. (\ref{equ:polarizabilties}), greatly simplifying the problem.


Since the particle is now modelled as a point dipole a set distance above the surface, the arbitrary integration volume in Eq. (\ref{equ:MSTsurf}) can be constricted to a single point around the point dipole and subsequent algebra yields the exact result \cite{Chaumet2009,Nieto-Vesperinas2010,Chen2011,Bliokh2014a}: 
\begin{equation}\label{equ:grad}
\begin{split}
\langle \mathbf{F} \rangle = \frac{1}{2} \mathbb{R} \bigg\{ &(\mathbf{\nabla} \otimes \widetilde{\mathbf{E}}) \mathbf{p}^* + \mu (\mathbf{\nabla} \otimes \widetilde{\mathbf{H}}) \mathbf{m}^* \\
& - \frac{k^4}{6 \pi \varepsilon c} (\mathbf{p} \times \mathbf{m}^*) \bigg\}
\end{split}
\end{equation}

\noindent where $c$ and $k$ are the speed of light and wavenumber of the medium enclosing the dipole, respectively. Note that the fields appearing in Eq. (\ref{equ:grad}) correspond to the background fields $\widetilde{\mathbf{E}}$ and $\widetilde{\mathbf{H}}$ and not the total fields. Also note that, with the condition that $\mathbf{p}$ and $\mathbf{m}$ are self-consistently generated, the force (\ref{equ:grad}) becomes linear with the background fields, enabling us to study the force from different contributions of the background field or even to its plane wave decomposition.

One could substitute Eqs. (\ref{equ:polarizabilties}) into the first two terms of Eq. (\ref{equ:grad}) and give rise to a range of force terms including a conservative scattering force, a nonconservative scattering force and a radiation pressure, among others \cite{Gao2017}. 
Throughout this paper, we will refrain from performing the substitution (\ref{equ:polarizabilties}) into (\ref{equ:grad}), for simplicity, so $\mathbf{p}$ and $\mathbf{m}$ must be calculated self-consistently.

Since $\widetilde{\mathbf{E}}$ and $\widetilde{\mathbf{H}}$ are the total fields excluding only the self-fields of the dipole, they of course include the illumination, but also, importantly, the backscattered fields (reflection of the self-fields of the dipole in the surface), sometimes called interaction fields. We can describe this with the substitution:
\begin{align}\label{equ:subFields}
\widetilde{\mathbf{E}} &\rightarrow \mathbf{E}^{\text{illum}} + \mathbf{E}^{\text{bs}} & \widetilde{\mathbf{H}} &\rightarrow \mathbf{H}^{\text{illum}} + \mathbf{H}^{\text{bs}}
\end{align}
\noindent where `illum' and `bs' relate to the illumination and backscattering contributions. 
The backscattered fields can also produce optical forces. This is a key point in this paper because the backscattering forces can be dominant when sufficiently close to the surface. We stress again that the backscattered fields depend on $\mathbf{p}$ and $\mathbf{m}$ and so must be calculated self-consistently such that Eqs. (\ref{equ:polarizabilties}) are fulfilled. This may involve multiple reflections. 

The third term in Eq. (\ref{equ:grad}) comes from the interference of the magnetic and electric dipole radiation and we refer to it as the radiative reaction term. An orthogonal and in-phase $\mathbf{p}$ and $\mathbf{m}$ can produce a far-field directionality due to the coherent interference of the two dipole radiations. This directionality has a perfect contrast in the case of a Huygens dipole \cite{Kerker1983,Picardi2017a,Chaumet2009,Gomez-Medina2012,Aunon2014}. The directional radiation produces a recoil force due to the conservation of linear momentum. 

The backscattered electromagnetic fields in Eq. (\ref{equ:subFields}) consist of contributions from each multipole of the particle. In this paper, we consider only the electric and magnetic dipoles, $\mathbf{E}^{\text{bs}} = \mathbf{E}_{e}^{\text{bs}} + \mathbf{E}_{m}^{\text{bs}}$ and $\mathbf{H}^{\text{bs}} = \mathbf{H}_{e}^{\text{bs}} + \mathbf{H}_{m}^{\text{bs}}$, but higher order multipoles may be added in a similar fashion. 

Many papers about optical tweezers on surfaces choose to omit the backscattered fields and just use the illumination fields as $\widetilde{\mathbf{E}}$ and $\widetilde{\mathbf{H}}$ in Eq. (\ref{equ:subFields}). This is done with the assumption that the forces due to the backscattered fields and any surface modes are negligible to the system. However, the addition of the backscattered fields in Eq. (\ref{equ:grad}) clearly shows the appearance of extra force terms which are not from the direct illumination. Several works do consider forces from the backscattered fields, and it has been shown that these forces can become very important to the motion of a particle close to a surface and so should not be neglected in general \cite{Rodriguez-Fortuno2014,Rodriguez-Fortuno2016,Rodriguez-Fortuno2018,Petrov2016,Paul2018,Wang2016,Wang2014}. We emphasize that any object that supports surface or guided modes can introduce non-trivial forces in this way due to a potentially directional dipole near-field. In other words, a recoil force is a completely general phenomenon that can occur for any source exhibiting near-field directionalities.

\section{\label{sec:ModeForce}Angular Force Spectrum}

We look to find an analytical expression of the near-field force for a dipole near a planar surface that clearly shows the recoil forces from guided or surface modes in a very general way. To do so, we analytically expand the fields $\widetilde{\mathbf{E}}$ and $\widetilde{\mathbf{H}}$ in Eq. (\ref{equ:grad}) into its angular spectrum.

By doing so, we arrive at the force angular spectrum \cite{Aunon2012,Aunon2013}, in which the force is written as an integral over the transverse wavevector, analogously to the field's angular spectrum \cite{Nieto-Vesperinas2006,Mandel1995,Novotny2006}. This type of expression lends itself very well to physical insight because the integrand can highlight exactly which angular components are contributing to a given force and with what magnitude. When plotting the integrand, a large area under the curve corresponds to a large force in the integrand and so resonances can play a key role in optical forces. 

Substituting Eq. (\ref{equ:subFields}) into Eq. (\ref{equ:grad}), the first and second terms can be arranged into forces caused exclusively by the illumination and by the backscattered fields, which combine with the radiative reaction force to give: $\langle \mathbf{F} \rangle = \langle \mathbf{F}^{\text{illum}} \rangle + \langle \mathbf{F}^{\text{bs}} \rangle + \langle \mathbf{F}^{\text{rr}} \rangle$. By substituting the angular spectrum of the backscattered fields into Eq. (\ref{equ:grad}), one can split the backscattering force into the near-field (evanescent) and far-field (propagating) components, $\langle \mathbf{F}^{\text{bs}} \rangle = \langle \mathbf{F}^{\text{bs}}_{\text{NF}} \rangle + \langle \mathbf{F}^{\text{bs}}_{\text{FF}} \rangle$. The resulting near-field force angular spectrum is given by (see Appendix A):
\PlaceText{123mm}{145mm}{$_{\text{(Fermi's golden rule)}}$}
\PlaceText{166mm}{145mm}{$_{\text{(Fermi's golden rule)}}$}
\begin{align}\label{equ:paperNewForce}
&\langle \mathbf{F}^{\text{bs}}_{\text{NF}} \rangle = \frac{1}{2} \mathbb{R} \bigg\{\iint_{\kappa} dk_x \, dk_y \, \frac{k^2}{8 \pi^2 \varepsilon}\frac{(-\mathbf{k})}{k_z}  \, e^{2i k_z h} \nonumber \\
&\Big(r_p \, \underbrace{\Big| \, \mathbf{p}^* \cdot \hat{\mathbf{E}}_p+ \mu \, \mathbf{m}^* \cdot \hat{\mathbf{H}}_p \, \Big|^2}_\text{p mode coupling} + \, r_s\, \underbrace{\Big| \, \mathbf{p}^* \cdot \hat{\mathbf{E}}_s + \mu \, \mathbf{m}^* \cdot \hat{\mathbf{H}}_s \, \Big|^2}_\text{s mode coupling}\Big) \bigg\} \nonumber \\
\end{align}
\begin{figure}[b]
\centering
\includegraphics[width=0.5\textwidth]{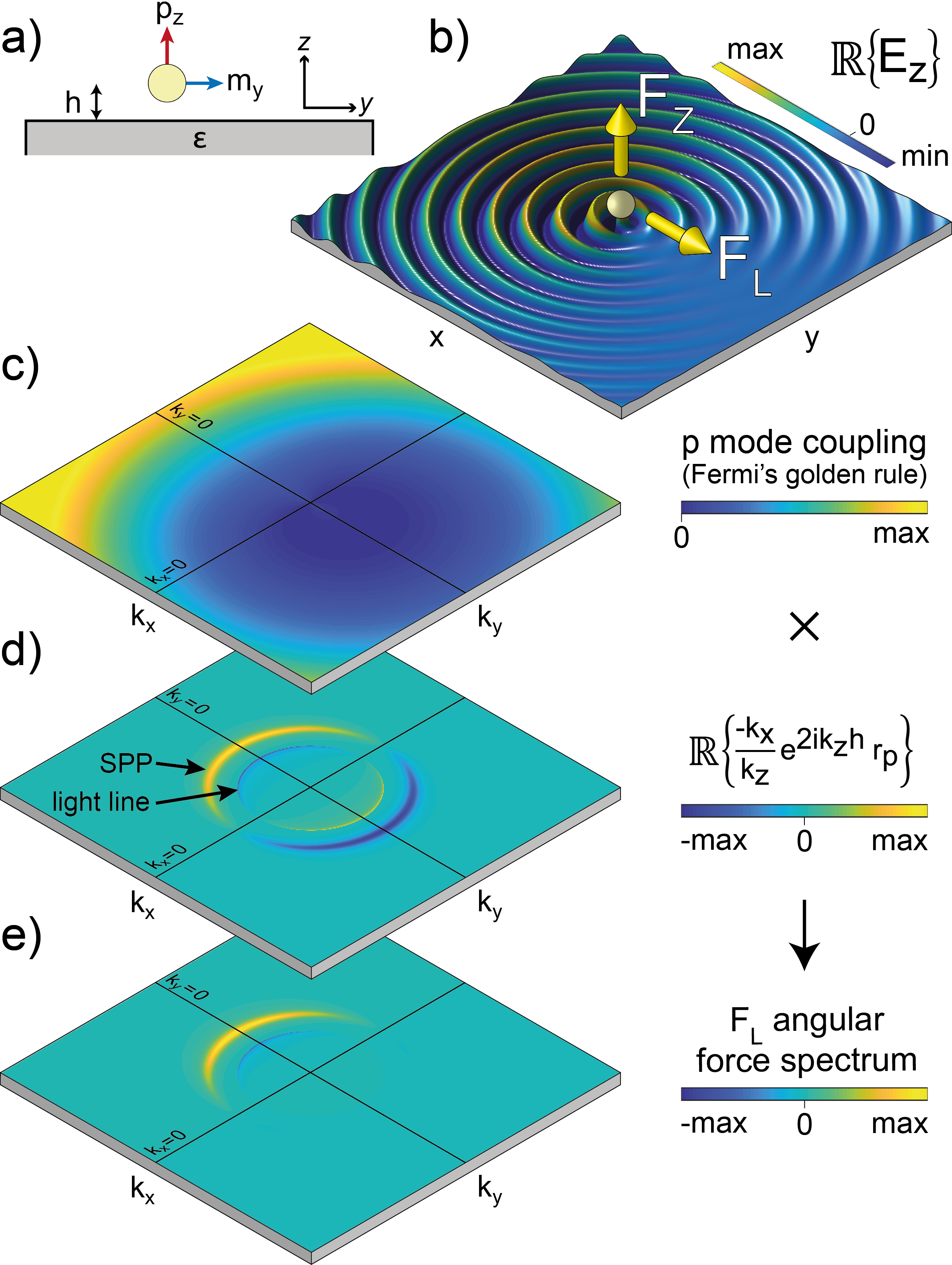}
\caption{(a) A schematic of the Huygens dipole $h$ above the surface of permittivity $\varepsilon$. (b) The electric field distribution of a p-polarized Huygens dipole $0.2 \lambda$ above an infinite planar surface of $\varepsilon_r = -2 + 0.2i$, illustrating the directional coupling. The yellow arrows depict the vertical and lateral forces. (c)-(e) Fourier plane spectra of (c) the p-polarized scattering from the Huygens dipole from Fermi's golden rule, (d) the $\mathbb{R}\{-k_x e^{2k_zh} r_p/k_{kz}\}$ term from Eq. (\ref{equ:paperNewForce}) and (e) the force along $x$, formed from multiplying (c) and (d). The SPP is the dominant contribution when the dipole is this close to the surface.}
\label{fig:AFSdiagram}
\end{figure}
\noindent where $\mathbf{k} = (k_x\hat{\mathbf{x}} + k_y\hat{\mathbf{y}} + k_z\hat{\mathbf{z}})$, $k_z = \sqrt[]{k^2-(k_x^2+k_y^2)}$, $\kappa$ symbolizes an integration over the domain where $\sqrt[]{k_x^2+k_y^2} >k$, associated to the near field, $h$ is the height of the dipole above the surface and $r_s$ and $r_p$ are the $k_x$ and $k_y$ dependent $s$ and $p$ polarized Fresnel reflection coefficients of the surface, respectively. $\hat{\mathbf{E}}_p$ and $\hat{\mathbf{H}}_p$ are the known normalized fields of any $p$-polarized evanescent wave,  given by $\hat{\mathbf{E}}_p$ = $\hat{\mathbf{e}}_p^+$ and $\hat{\mathbf{H}}_p$ = $(1/\eta)\,\hat{\mathbf{e}}_s$. Likewise, $\hat{\mathbf{E}}_s$ and $\hat{\mathbf{H}}_s$ are the normalized fields of any $s$-polarized evanescent wave, given by $\hat{\mathbf{E}}_s = \hat{\mathbf{e}}_s$ and $\hat{\mathbf{H}}_s = -(1/\eta)\,\hat{\mathbf{e}}_p^+$. We use the $s$ and $p$ polarization basis vectors defined in Ref. \cite{Rotenberg2012, Picardi2017} as $\hat{\mathbf{e}}_s = (k_x^2+k_y^2)^{-\frac{1}{2}}(-k_y \hat{\mathbf{x}} + k_x \hat{\mathbf{y}})$ and $\hat{\mathbf{e}}_p^\pm = \hat{\mathbf{e}}_s \times \frac{\mathbf{k}^\pm}{k}$, where $\mathbf{k}^\pm = (k_x \hat{\mathbf{x}} + k_y \hat{\mathbf{y}} \pm k_z \hat{\mathbf{z}})$ and $\pm$ refers to propagation in the $+\hat{\mathbf{z}}$ or $-\hat{\mathbf{z}}$ direction. These basis vectors correspond to the well known unit vectors of the azimuthal and polar angles in spherical coordinates when $\mathbf{k}$ is real. Note that $r_s$, $r_p$, $\hat{\mathbf{e}}_s$ and $\hat{\mathbf{e}}_p^\pm$ are complex and functions of $k_x$ and $k_y$.
While $\langle \mathbf{F}^{\text{bs}}_{\text{NF}} \rangle$ is only suitable for the near-field domain, $\langle \mathbf{F}^{\text{bs}} \rangle \approx \langle \mathbf{F}^{\text{bs}}_{\text{NF}} \rangle$ if suitably close to the surface (subwavelength distance) due to the evanescent contributions dominating the integrals. A less compact expression valid for the whole transverse wavevector plane is provided in Appendix A. 

Fermi's golden rule describes the coupling efficiency between a dipole and waveguide \cite{Picardi2017a,VanMechelen2016,Espinosa-Soria2016} and can be expressed as $|A_m|^2 \propto |\mathbf{p}^* \cdot \mathbf{E}_m + \mu \, \mathbf{m}^* \cdot \mathbf{H}_m|^2$, where $|A_m|^2$ is the coupling efficiency and $\mathbf{E}_m$ and $\mathbf{H}_m$ are the fields of the mode being considered. By considering only $p$ polarized modes with Fermi's golden rule ($\mathbf{E}_m \rightarrow \mathbf{E}_p$ and $\mathbf{H}_m \rightarrow \mathbf{H}_p$), it is clear that the normalized coupling efficiency of $p$ polarized modes is apparent in Eq. (\ref{equ:paperNewForce}), weighted by $r_p$. The same is true for $s$ polarized modes. It is both surprising and remarkable that Fermi's golden rule appears in such an elegant way inside the apparently unrelated angular force spectrum.

Fig. \ref{fig:AFSdiagram} demonstrates how the different terms combine to produce the net lateral force: Fig. \ref{fig:AFSdiagram}c shows the angular representation of the $p$ polarized coupling efficiency, determined by Fermi's golden rule as written in Eq. (\ref{equ:paperNewForce}), and depends exclusively on the dipole moments $\mathbf{p}$ and $\mathbf{m}$ and their near and far field directionalities \cite{Picardi2017a,Picardi2017}. Notice that Fig 1c is completely independent of the surface being considered. The clear asymmetry in Fig. \ref{fig:AFSdiagram}c represents the dipolar directionality, and is ultimately responsible for the existence of lateral forces. In contrast, Fig. \ref{fig:AFSdiagram}d is independent of the polarization of the dipolar source and its directionality, but it contains the Fresnel reflection coefficient, which represents the optical response of the surface (in this case a metallic surface supporting surface plasmons). Fig. \ref{fig:AFSdiagram}d also includes the dependence of the force on the distance between the dipole and the surface, and the $\frac{-\mathbf{k}}{k_z}$ factor in Eq. (\ref{equ:paperNewForce}).
The $-\mathbf{k}$ factor means the force is anti-parallel to the propagation direction of the excited mode. This is physically intuitive, as a strong mode coupling to a waveguide along a given direction will incite a strong recoil force in the opposite direction, as dictated by conservation of momentum. 

The product of Fig 1c, dependent only on the dipole, with Fig 1d, dependent on the surface, the distance and the force's direction, produces the final force angular spectrum shown in Fig 1e, which represents the combined effect of the dipole and the surface and whose integration results in the total force. This visual representation clearly unveils the physical origins of the force from different contributions.

Even though each evanescent component of the backscattered field produces an associated recoil force, the total force can be zero due to the different components canceling each other when the amplitudes of guided modes in different directions (calculated via Fermi's golden rule) are balanced.  Only when there is an imbalance in the coupling to guided modes (known as near-field directionality \cite{Picardi2017a}), such as in Fig. \ref{fig:AFSdiagram}, can we have a non zero net recoil force. Another way to obtain non zero net recoil forces (even for non-directional dipoles) would be to use a surface with angular-dependent reflection coefficients, $r_s$ and $r_p$, which can be achieved in magneto-optical materials or metals by applying static magnetic fields \cite{Yu2008, Chettiar2014}.

An interesting observation can be made about the $\frac{\mathbf{k}}{k_z}$ prefactor of the force angular spectrum. While the $\hat{\mathbf{z}}$ component is always equal to unity, the transverse components become imaginary in the near-field regime, switching the force to depend on the imaginary part of the reflection coefficients. This is in contrast to the far-field regime where all components of the force are functions of the real parts of $r_s$ and $r_p$. 

\section{\label{sec:NumInv}Numerical investigation}

\begin{figure}[b]
\centering
\includegraphics[width=0.5\textwidth]{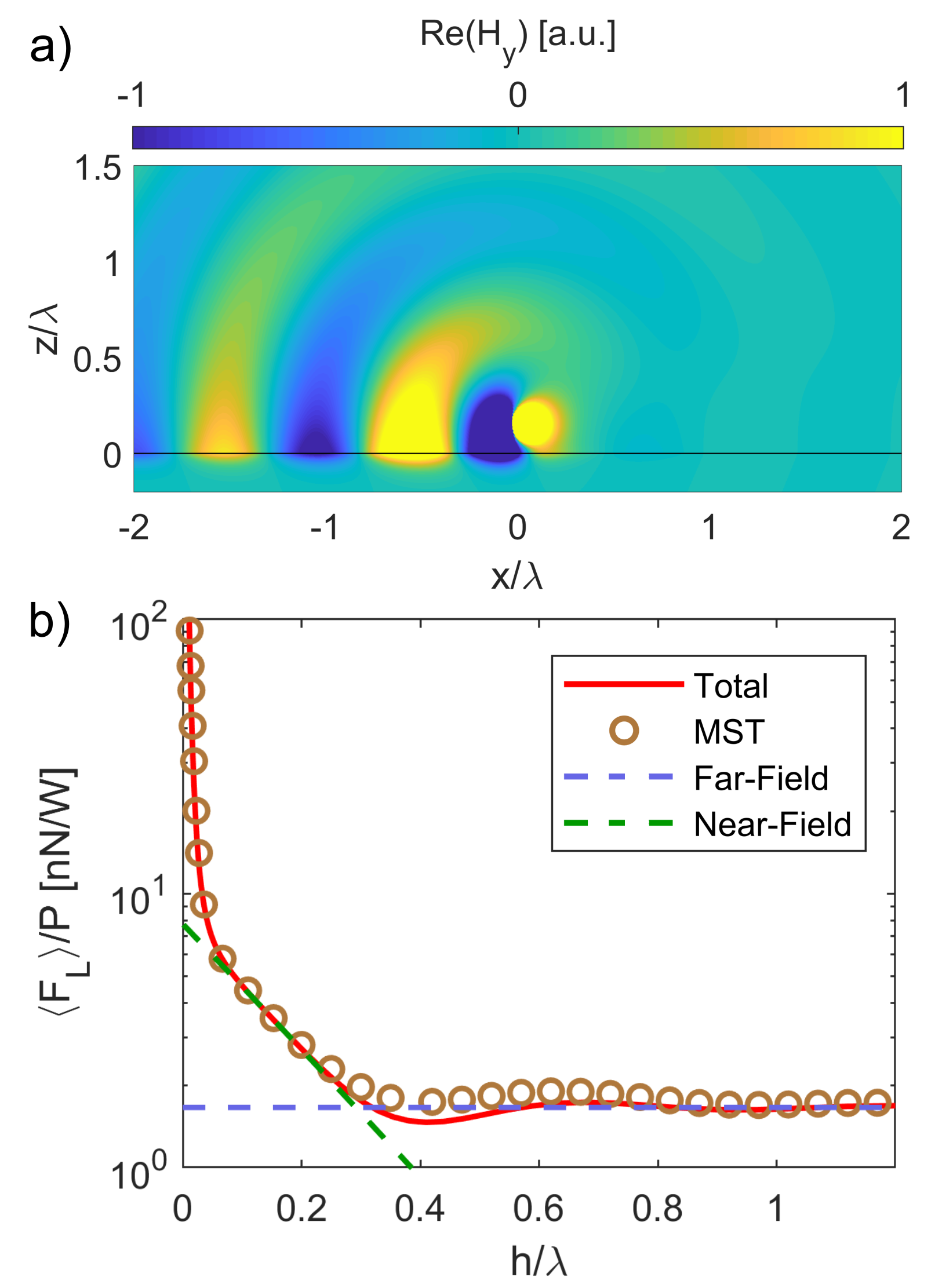}
\caption{(a) Magnetic field strength plotted for an optimized p-polarized Huygens dipole at a height $h = 0.15 \lambda$ above a metallic plane at $z = 0$ of $\varepsilon_r = -12 + 3i$ and $\mu_r = 1$. The directional SPP is clearly visible around the surface at $z/\lambda = 0$.
(b) The time-averaged lateral force on the same Huygens dipole with respect to $h$ over the same surface, normalized by the scatting power. Total refers to the force calculated via (Eq. \ref{equ:grad}) and MST refers to the force calculated via the Maxwell stress tensor (Eq. \ref{equ:MSTsurf}). The propagating far-field force and SPP recoil force are shown with lines.}
\label{fig:FxHuy}
\end{figure}
To numerically illustrate the significance of these backscattering forces, we explore the simple example of a Rayleigh particle with electric and magnetic dipole moments, $\mathbf{p}$ and $\mathbf{m}$, that form a Huygens dipole and place it near a homogeneous flat surface. We model only the radiating dipole and its fields reflected from the surface, $\mathbf{E}^{\text{bs}}$ and $\mathbf{H}^{\text{bs}}$, because we wish to highlight the particle-waveguide interactions and their effects on the optical force. Including the illuminating fields, $\mathbf{E}^{\text{illum}}$ and $\mathbf{H}^{\text{illum}}$, would yield no new physical insights and so are not included in the fields of Eq. (\ref{equ:grad}) for this example. Throughout this section, we refer to gradient forces when describing the forces from the gradient of the reflected fields, which correspond to the first two terms of Eq. (\ref{equ:grad}). The dipole's fields are reflected by the surface in accordance with the Fresnel reflection coefficients, $r_p$ and $r_s$, and we employ the angular representation to compute the electromagnetic fields \cite{Picardi2017}. 

Fig. \ref{fig:FxHuy} shows the lateral force felt by an optimized Huygens dipole ($m_y= k_{spp} \, c \, p_z$, where $k_{spp} = \sqrt[]{\frac{\varepsilon_r}{1+\varepsilon_r}}$) \cite{Picardi2017a} in close proximity to a metallic plane at $z = 0$ and was calculated by inserting the backscattered fields into Eq. (\ref{equ:grad}). The Huygens dipole is an excellent example, because it possesses a far-field directionality whose recoil force $\mathbf{F}^{\text{rr}}$ is given by the radiative reaction term in Eq. (\ref{equ:grad}), while at the same time, as shown in \cite{Picardi2017a}, it also exhibits a near field directionality, which results in lateral near-field recoil forces, $\mathbf{F}^{\text{bs}}$. Thus, this system is ideal in order to compare the relative magnitude of the commonly used far-field radiative reaction force $\mathbf{F}^{\text{rr}}$ and the typically neglected gradient forces caused by the backscattered fields $\mathbf{F}^{\text{bs}}$. It is already understood that the far-field radiative reaction force $\mathbf{F}^{\text{rr}}$ is significant enough to overcome the illumination gradient force $\mathbf{F}^{\text{illum}}$ \cite{Chen2011,Novitsky2011}, so any forces stronger than this are also significant. The results were confirmed through use of the MST (Eq. \ref{equ:MSTsurf}). 

It is clear that the plot has three distinct regions. 
The first region at $h/\lambda>0.3$ has a decaying sinusoidal shape with a period of $\lambda/2$ around a non-zero lateral force. The non-zero equilibrium point is the far-field recoil force of the Huygens dipole $\mathbf{F}^{\text{rr}}$, corresponding to the radiative reaction term in Eq. (\ref{equ:grad}). The weak oscillation with $\lambda/2$ periodicity comes from the gradient force of the reflected far field of the dipole $\mathbf{F}_{\text{FF}}^\text{bs}$ which can be neglected at these heights, as is usually done.

\begin{figure}[b]
\centering
\includegraphics[width=0.5\textwidth]{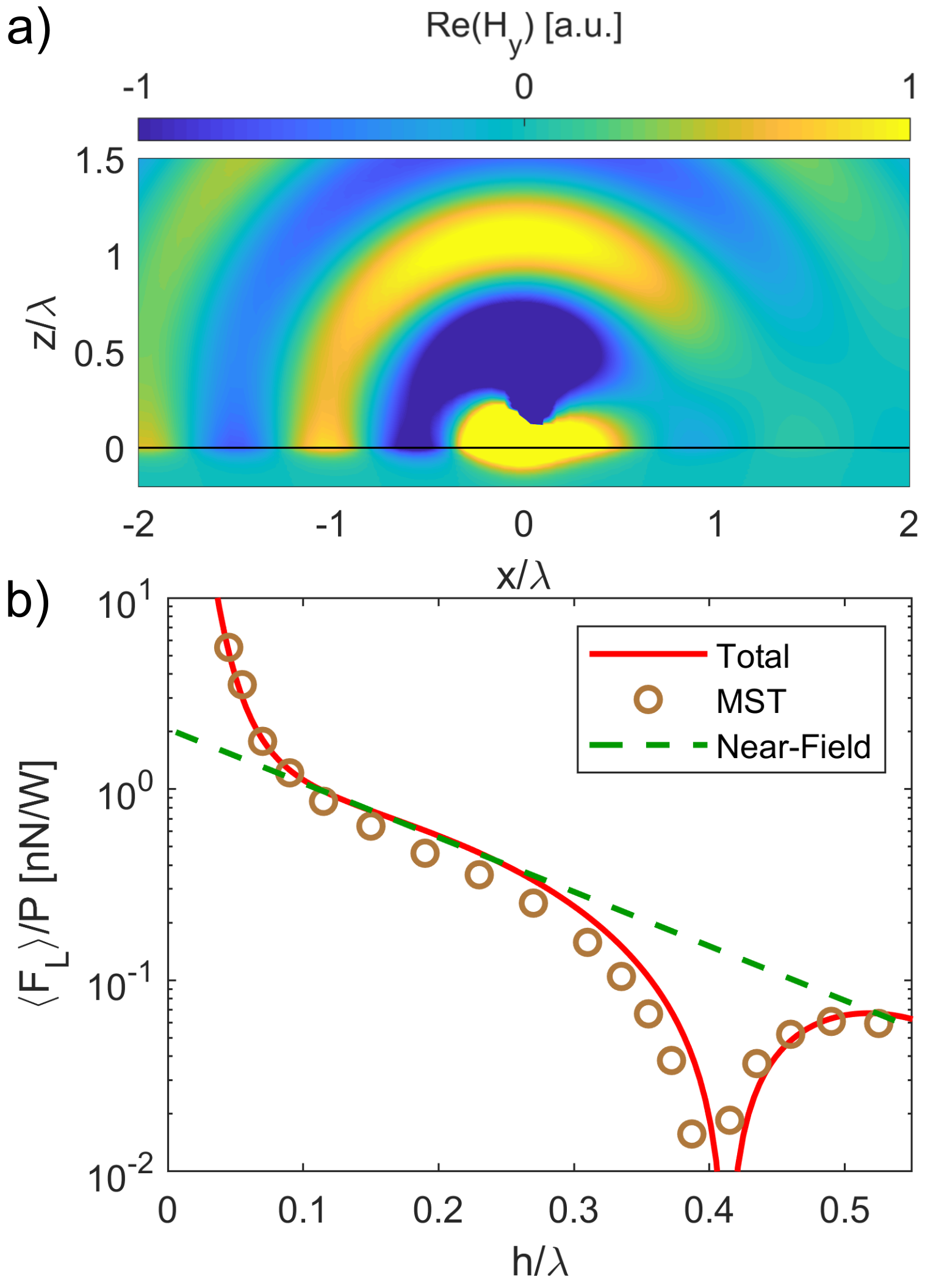}
\caption{(a) Magnetic field strength plotted for an optimized elliptical dipole at $h=0.15\lambda$ above the a surface of $\varepsilon_r =  -12 + 3i$. The directional SPP is again clearly visible around the surface at $z/\lambda = 0$. (b) The modulus of the normalized lateral force on the same dipole with respect to $h$ above the same surface, normalized by the scatting power. Total refers to the force calculated via (Eq. \ref{equ:grad}) and MST refers to the force calculated via the Maxwell stress tensor (Eq. \ref{equ:MSTsurf}). Elliptical dipoles have no far-field directionality when far from a surface.}
\label{fig:FxCirc}
\end{figure}

The other two regions are dominated by the force due to backscattered fields, $\mathbf{F}^{\text{bs}}_{\text{NF}}$, showing that they cannot be neglected when the surface is within the near-fields of the dipole.
The second region at $h/\lambda<0.1$ is in the quasistatic limit where the integral in the angular representation is dominated by large transverse wavevectors $k_t$ and results in the dipole `feeling' an image dipole placed at a distance $h$ into the surface \cite{Rodriguez-Fortuno2014}. The resultant force is asymptotic but in a realistic system with a particle of finite size, this region may be unreachable. 

The region of interest at $0.1 < h/\lambda < 0.3$ a main focus of this paper.
A mode pertaining to the surface has been excited by the near-field directionality of the dipole and a near-field recoil force $\mathbf{F}_{\text{NF}}^{\text{bs}}$ has been produced. The force varies exponentially with the separation distance from the surface because of the decaying nature of the reflected evanescent waves, as predicted by Eq. (\ref{equ:paperNewForce}). In this particular example, the excited mode is a surface plasmon polariton (SPP) on the metallic surface (see Fig. \ref{fig:FxHuy}) and it can contribute overwhelmingly  to the total force on the dipole, via the near-field recoil force $\mathbf{F}^{\text{bs}}_{\text{NF}}$. For a dielectric waveguide, a similar force would appear because of any excited guided modes. 
The lateral force exists due to the near-field directionality of the Huygens dipole, as shown clearly in the fields of Fig. \ref{fig:FxHuy}, and this directionality is independent of the surface being used. As shown by Eq. (\ref{equ:paperNewForce}), the surface will only determine the strength of the force via the guided modes which appear as peaks in the Fresnel reflection coefficient, but the direction and existence of the lateral force is ultimately stemming from the p-mode coupling near-field directionality of the dipole itself.

This behavior is very general and applies to all dipoles. For example, Fig. \ref{fig:FxCirc} shows the near-field lateral force $\mathbf{F}_{\text{NF}}^{\text{bs}}$ from an optimized \cite{Picardi2017a} elliptical dipole $\mathbf{p} = (k_{spp}, 0, -i \, \sqrt{k_{spp}^2-1})$. The same dominance of the SPP in the $0.1<h/\lambda<0.3$ region demonstrates that this physics is not unique to the Huygens. An elliptical dipole is known to have no far field directionality in vacuum and therefore no far-field force $\mathbf{F}^{\text{rr}}$. This leads to the $h/\lambda>0.3$ region oscillating about $\langle F_L\rangle = 0$ rather than the non-zero far-field force $\mathbf{F}^{\text{rr}}$ of Fig. \ref{fig:FxHuy}. The oscillation itself can be said to be the far-field directionality that the elliptical dipole gains when above a metallic surface \cite{Xi2013}. The near-field directionality and therefore near-field forces can be directly controlled by the polarization of the dipole, which experimentally can be controlled via illuminating polarization. For both examples and any other dipoles, the directional near-field forces can become much larger than the far-field forces and so cannot be neglected. 

The mode recoil forces are a completely general result applying to any directional guided or surface mode that exists on any planar structure. The forces appear as quasistatic, guided/surface mode and far-field forces. The same principles of recoil forces due to near fields can be applied in arbitrary non-planar geometries by using directly Eqs. (\ref{equ:MSTsurf}) or (\ref{equ:grad}), but without such easy analytical form as in the planar case.

\section{\label{sec:Concl}Conclusions}

In this paper, we discussed the significance of near-field interactions on optical forces and the underlying principles of symmetry and conservation of momentum behind them. We showed a link between the near-field recoil force on coupling dipole and the excitation amplitude of the surface mode which is determined by Fermi's golden rule. 

We have compared the forces upon a general electromagnetic dipole in the limit where a surface is placed close to the particle, and demonstrate near-field interactions begin to dominate the forces on the particle. The Huygens dipole and an elliptically polarized electric dipole are used as examples that exhibit near-field directionality and show that the resultant backscattered fields are crucial for accurately calculating the forces on the dipoles. This model is highly applicable to a wide range of systems where a Rayleigh nanoparticle or nanoantenna is in close proximity to a waveguide and exhibits sharp electric and/or magnetic dipole resonances under illumination. Any nanomanipulation in this regime must take the backscattered near-fields into consideration.

\section{\label{sec:Acknow}Acknowledgements}

This work was supported by European Research Council Starting Grant ERC-2016-STG-714151-PSINFONI, EPSRC (UK), ERC iCOMM project (789340), the Royal Society and the Wolfson Foundation. All the data supporting this research are provided in full in the results section and appendices.

\setcounter{equation}{0}
\renewcommand{\theequation}{A\arabic{equation}}

\appendix
\section{\label{sec:AppA}APPENDIX A: FORCE ANGULAR SPECTRUM DERIVATION}

To describe the force on a general magnetodielectric particle near a planar surface, we can construct the force angular spectrum for a given frequency. We begin with the angular spectrum of the electric field \cite{Nieto-Vesperinas2006,Mandel1995,Novotny2006}. 
\begin{align}\label{equ:angularspectrum}
\mathbf{E}(x,y,z) = \iint\mathbf{E}(k_x,k_y,z) \, e^{i(k_x\,x + \, k_y \, y)}\, dk_x \, dk_y
\end{align}
\noindent The magnetic field can be expressed in a similar manner. Let's now consider only the force caused by the backscattered or reflected fields, $\langle \mathbf{F}^{\text{bs}} \rangle$. 
The reflected field contributions from the electric and magnetic dipoles are combined: $\mathbf{E}_e^{\text{bs}} + \mathbf{E}_m^{\text{bs}} = \mathbf{E}_{\text{total}}^{\text{bs}}$. We then substitute (\ref{equ:angularspectrum}) into the gradient terms of Eq. (\ref{equ:grad}) 
and evaluate the gradient operator as equivalent to a factor $i\, \mathbf{k}$ when working with angular spectra. After the substitution of the mathematical identity ($i\, \mathbf{k}\, \otimes \mathbf{E}) \mathbf{p}^* = i\, \mathbf{k} (\mathbf{E} \cdot \mathbf{p}^*$), and similarly for the magnetic dipole term, we arrive at:
\begin{align}\label{equ:Fproptoike}
\langle \mathbf{F}^{\text{bs}} \rangle = \frac{1}{2} \mathbb{R} \bigg\{\iint i\,\mathbf{k} \, \Big(\mathbf{p}^* \cdot \mathbf{E}_{\text{total}}^{\text{bs}} + \mu \, \mathbf{m}^* \cdot \mathbf{H}_{\text{total}}^{\text{bs}}\Big) \, dk_x \, dk_y\bigg\}
\end{align}

\noindent where $\mathbf{k} = (k_x\,\hat{\mathbf{x}} + k_y\,\hat{\mathbf{y}} + k_z\,\hat{\mathbf{z}})$ and the integrals are conducted from $-\infty$ to $+\infty$ in  $k_x$ and $k_y$. 
The electromagnetic fields can be expanded in the angular representation with the polarization vectors discussed in Ref. \cite{Rotenberg2012, Picardi2017}. For a dipole above a planar surface at $z = 0$, the reflected dipole fields at the position of the dipole $\mathbf{r} = h\, \hat{\mathbf{z}}$ are given by the angular representation \cite{Picardi2017}
\begin{align}\label{equ:ReflFields}
\mathbf{E}&_e^{\text{bs}}(k_x,k_y,h) \nonumber \\
&=  \frac{i \, k^2}{8 \pi^2 \, \varepsilon \, k_z} \big[r_p(\hat{\mathbf{e}}_p^- \cdot \mathbf{p})\hat{\mathbf{e}}_p^+ + r_s(\hat{\mathbf{e}}_s \cdot \mathbf{p})\hat{\mathbf{e}}_s\big] e^{2 i k_z h} \nonumber \\
\mathbf{E}&_m^{\text{bs}}(k_x,k_y,h) \nonumber \\
&=  \frac{i \, k^2}{8 \pi^2  \,\varepsilon \, k_z c} \big[r_p(\hat{\mathbf{e}}_s \cdot \mathbf{m})\hat{\mathbf{e}}_p^+ - r_s(\hat{\mathbf{e}}_p^- \cdot \mathbf{m})\hat{\mathbf{e}}_s \big] e^{2 i k_z h} \nonumber \\
\mu\, \mathbf{H}&_e^{\text{bs}}(k_x,k_y,h) \nonumber \\
&=  \frac{i \, k^2}{8 \pi^2  \,\varepsilon \, k_z c} \big[r_p(\hat{\mathbf{e}}_p^- \cdot \mathbf{p})\hat{\mathbf{e}}_s - r_s(\hat{\mathbf{e}}_s \cdot \mathbf{p})\hat{\mathbf{e}}_p^+ \big] e^{2 i k_z h} \nonumber \\
\mu\, \mathbf{H}&_m^{\text{bs}}(k_x,k_y,h) \nonumber \\
&=  \frac{i \, k^2}{8 \pi^2  \,\varepsilon \, k_z c^2} \big[r_p(\hat{\mathbf{e}}_s \cdot \mathbf{m})\hat{\mathbf{e}}_s + r_s(\hat{\mathbf{e}}_p^- \cdot \mathbf{m})\hat{\mathbf{e}}_p^+ \big] e^{2 i k_z h}\nonumber\\
\end{align}
\noindent where $k$ is the wavenumber of the medium enclosing the dipole, $\varepsilon$ is the permittivity and $r_s = r_s(k_x,k_y)$ and $r_p = r_p(k_x,k_y)$ are the $s$ and $p$ polarized Fresnel reflection coefficients, respectively. Note that these reflected fields can (and will) involve any number of multiple reflections so long as $\mathbf{p}$ and $\mathbf{m}$ are solved self-consistently with Eqs. (\ref{equ:polarizabilties}).

The $\mathbf{e}_s$ and $\mathbf{e}_p^\pm$ vectors form a set of orthogonal basis vectors ($\mathbf{e}_s \cdot \mathbf{e}_p^\pm = 0$) and are applicable to both propagating and evanescent waves.

After substituting (\ref{equ:ReflFields}) into (\ref{equ:Fproptoike}), the integrand of (\ref{equ:Fproptoike}) can be split into $r_s$ and $r_p$ contributions and then factorized. This leads directly to the force angular spectrum of the dipole near any planar surface:
\begin{align}
\begin{split}\label{equ:generalforce}
\langle \mathbf{F}^{\text{bs}} \rangle &= -\frac{1}{2} \mathbb{R} \bigg\{\iint \, dk_x \, dk_y \, e^{2i k_z h} \frac{k^2 \, \mathbf{k}}{8 \pi^2 \varepsilon k_z}  \\
&\Big[ r_p \, \Big(\mathbf{p}^* \cdot \hat{\mathbf{e}}_p^+ + \frac{\mathbf{m}^*}{c} \cdot \hat{\mathbf{e}}_s \, \Big)
\Big(\mathbf{p} \cdot \hat{\mathbf{e}}_p^{-} + \frac{\mathbf{m}}{c} \cdot \hat{\mathbf{e}}_s \, \Big)\\
&+ \, r_s\, \Big( \, \mathbf{p}^* \cdot \hat{\mathbf{e}}_s - \frac{\mathbf{m}^*}{c} \cdot \hat{\mathbf{e}}_p^+ \, \Big)
\Big( \, \mathbf{p} \cdot \hat{\mathbf{e}}_s - \frac{\mathbf{m}}{c} \cdot \hat{\mathbf{e}}_p^{-} \, \Big)\Big]\bigg\}\\
\end{split}
\end{align}

Eq. (\ref{equ:generalforce}) integrates over the whole $k_x$, $k_y$ plane. This plane can be decomposed into the far-field region $\sqrt[]{k_x^2+k_y^2} <k$ and the near-field region $\sqrt[]{k_x^2+k_y^2} >k$, corresponding to propagating and evanescent components, respectively. In the near-field region, $\kappa$, (\ref{equ:generalforce}) can be greatly simplified because $k_z$ is purely imaginary, leading to the near-field properties of the unit vectors of $\hat{\mathbf{e}}_p^\pm = \hat{\mathbf{e}}_p^{\mp*}$ and $\hat{\mathbf{e}}_s = \hat{\mathbf{e}}_s^{*}$. Applying these to Eq. (\ref{equ:generalforce}) gives:
\begin{align}\label{equ:guidetoFGR}
\langle \mathbf{F}_{\text{NF}}^{\text{bs}} \rangle &= -\frac{1}{2} \mathbb{R} \bigg\{\iint_{\kappa} \, dk_x \, dk_y \, e^{2i k_z h} \frac{k^2 \, \mathbf{k}}{8 \pi^2 \varepsilon k_z}  \nonumber\\
\Big[ r_p& \, \Big(\mathbf{p}^* \cdot \hat{\mathbf{e}}_p^+ + \frac{\mathbf{m}^*}{c} \cdot \hat{\mathbf{e}}_s \, \Big)\nonumber
\Big(\mathbf{p} \cdot \hat{\mathbf{e}}_p^{+*} + \frac{\mathbf{m}}{c} \cdot \hat{\mathbf{e}}_s^* \, \Big)\nonumber\\
+ \, r_s& \, \Big( \, \mathbf{p}^* \cdot \hat{\mathbf{e}}_s - \frac{\mathbf{m}^*}{c} \cdot \hat{\mathbf{e}}_p^+ \, \Big)\nonumber
\Big( \, \mathbf{p} \cdot \hat{\mathbf{e}}_s^* - \frac{\mathbf{m}}{c} \cdot \hat{\mathbf{e}}_p^{+*} \, \Big)\Big]\bigg\}\nonumber\\
\nonumber\\
&= -\frac{1}{2} \mathbb{R} \bigg\{\iint_{\kappa} dk_x \, dk_y \, \frac{k^2}{8 \pi^2 \varepsilon}\frac{\mathbf{k}}{k_z}  \, e^{2i k_z h} \nonumber\\
\Big(r_p \, \Big| \, \mathbf{p}^* \cdot \, &\hat{\mathbf{e}}_p^++ \frac{\mathbf{m}^*}{c} \cdot \hat{\mathbf{e}}_s \, \Big|^2 + \, r_s\, \Big| \, \mathbf{p}^* \cdot \hat{\mathbf{e}}_s - \frac{\mathbf{m}^*}{c} \cdot \hat{\mathbf{e}}_p^+ \, \Big|^2\Big) \bigg\}
\end{align}

\noindent where we can identify the normalized mode field vectors $\hat{\mathbf{E}}_s = \hat{\mathbf{e}}_s$, $\hat{\mathbf{E}}_p = \hat{\mathbf{e}}_p^\pm$, $\hat{\mathbf{H}}_s = -(1/\eta)\,\hat{\mathbf{e}}_p^\pm$ and $\hat{\mathbf{H}}_p = (1/\eta)\,\hat{\mathbf{e}}_s$, where $\eta = \sqrt[]{\mu/\varepsilon}$. By substituting these vectors into Eq. (\ref{equ:guidetoFGR}), we arrive at Eq. (\ref{equ:paperNewForce}). $\hat{\mathbf{e}}_p^+$ appears in Eq. (\ref{equ:guidetoFGR}) because the plane is beneath the dipole and so the radiation from the surface or guided mode is propagating towards the dipole in the positive $\hat{\mathbf{z}}$ direction. 

Eq. (\ref{equ:guidetoFGR}) can be displayed in an alternative form where terms are split according to dipole interactions rather than polarizations by making use of the complex conjugate identity:
\begin{align*}
|A+B|^2 = |A|^2 + |B|^2 + 2 \, \mathbb{R} \{ A B^* \}
\end{align*}
\noindent Some readers may prefer this alternative expression. 
\begin{align*}\label{equ:forcesplit}
\langle \mathbf{F}_{\text{NF}}^{\text{bs}} \rangle &= -\frac{1}{2} \mathbb{R} \bigg\{\iint_{\kappa} \mathbf{\gamma} \, \big[f_e + f_m + f_{em}\big] \, dk_x \, dk_y\bigg\},\\
\mathbf{\gamma} &=  \, \frac{k^2 \, \mathbf{k}}{8 \pi^2 \varepsilon k_z}  \, e^{2i k_z h}\nonumber\\
f_{e} &=  r_p \, |\mathbf{p}^* \cdot \hat{\mathbf{e}}_p^+|^2 +r_s \, |\mathbf{p}^* \cdot \hat{\mathbf{e}}_s|^2\nonumber\\
f_{m} &= r_p \, \Big|\frac{\mathbf{m}^*}{c} \cdot \hat{\mathbf{e}}_s\Big|^2 + r_s \, \Big|\frac{\mathbf{m}^*}{c} \cdot \hat{\mathbf{e}}_p^+\Big|^2\nonumber\\
f_{em} &= 2 \, r_p \, \mathbb{R}\Big\{(\mathbf{p}^* \cdot \hat{\mathbf{e}}_p^+)\Big(\frac{\mathbf{m}}{c} \cdot \hat{\mathbf{e}}_s\Big)\Big\}\nonumber\\
&\hspace{0.5cm}- 2 \, r_s \, \mathbb{R}\Big\{(\mathbf{p}^* \cdot \hat{\mathbf{e}}_s)\Big(\frac{\mathbf{m}}{c} \cdot \hat{\mathbf{e}}_p^{-}\Big)\Big\} \nonumber
\end{align*}

\section{\label{sec:AppB}APPENDIX B: CROSS POLARISATION}
\setcounter{equation}{0}
\renewcommand{\theequation}{B\arabic{equation}}

The Fresnel reflection coefficients can be generalized for surfaces that can convert polarizations of light upon reflection.
\begin{equation}\label{equ:crosspoldef}
\begin{pmatrix}
E^{\text{ref}}_p \\
E^{\text{ref}}_s
\end{pmatrix}
=
\begin{pmatrix}
r_{pp} & r_{ps} \\
r_{sp} & r_{ss}
\end{pmatrix}
\begin{pmatrix}
E^{\text{inc}}_p \\
E^{\text{inc}}_s
\end{pmatrix}
\end{equation}
\noindent where `inc' and `ref' refer to any field incident on the surface $\mathbf{E}^\text{inc} = (E_p^\text{inc} \, \hat{\mathbf{e}}_p^- + E_s^\text{inc} \, \hat{\mathbf{e}}_s) \, e^{i(k_x \, x +k_y \, y - k_z \, z)}$ and the subsequent reflection $\mathbf{E}^\text{ref} = (E_p^\text{ref} \, \hat{\mathbf{e}}_p^+ + E_s^\text{ref} \, \hat{\mathbf{e}}_s) \, e^{i(k_x \, x +k_y \, y + k_z \, z)}$, respectively. The angular spectrum expressions of Ref. \cite{Picardi2017} can be generalized for the $r_{sp}$ and $r_{ps}$ cross polarization terms. Just as before, applying the gradient terms of Eq. (\ref{equ:grad}) will produce an angular spectrum of the form:
\begin{align}
\begin{split}\label{equ:crosspolraw}
\langle \mathbf{F}^{\text{bs}} \rangle &= -\frac{1}{2} \mathbb{R} \bigg\{\iint \, dk_x \, dk_y \, e^{2i k_z h} \frac{k^2 \, \mathbf{k}}{8 \pi^2 \varepsilon k_z}  \\
&\Big[ r_{pp} \, \Big(\mathbf{p}^* \cdot \hat{\mathbf{e}}_p^+ + \frac{\mathbf{m}^*}{c} \cdot \hat{\mathbf{e}}_s \, \Big)
\Big(\mathbf{p} \cdot \hat{\mathbf{e}}_p^{-} + \frac{\mathbf{m}}{c} \cdot \hat{\mathbf{e}}_s \, \Big)\\
&+ \, r_{ps}\, \Big( \, \mathbf{p}^* \cdot \hat{\mathbf{e}}_p^+ + \frac{\mathbf{m}^*}{c} \cdot \hat{\mathbf{e}}_s \, \Big)
\Big( \, \mathbf{p} \cdot \hat{\mathbf{e}}_s - \frac{\mathbf{m}}{c} \cdot \hat{\mathbf{e}}_p^{-} \, \Big)\\
&+ \, r_{sp}\, \Big( \, \mathbf{p}^* \cdot \hat{\mathbf{e}}_s - \frac{\mathbf{m}^*}{c} \cdot \hat{\mathbf{e}}_p^+ \, \Big)
\Big( \, \mathbf{p} \cdot \hat{\mathbf{e}}_p^- + \frac{\mathbf{m}}{c} \cdot \hat{\mathbf{e}}_s \, \Big)\\
&+ \, r_{ss}\, \Big( \, \mathbf{p}^* \cdot \hat{\mathbf{e}}_s - \frac{\mathbf{m}^*}{c} \cdot \hat{\mathbf{e}}_p^+ \, \Big)
\Big( \, \mathbf{p} \cdot \hat{\mathbf{e}}_s - \frac{\mathbf{m}}{c} \cdot \hat{\mathbf{e}}_p^{-} \, \Big)\Big]\bigg\}\\
\end{split}
\end{align}

\noindent Eq. (\ref{equ:crosspolraw}) can be written in the same form as Eq. (\ref{equ:paperNewForce}) by substituting in the previously mentioned normalized mode field vectors to produce:
\begin{align*}
\begin{split}\label{equ:crosspolfields}
\langle \mathbf{F}_{\text{NF}}^{\text{bs}} \rangle &= -\frac{1}{2} \mathbb{R} \bigg\{\iint_{\kappa} \, dk_x \, dk_y \, e^{2i k_z h} \frac{k^2 \, \mathbf{k}}{8 \pi^2 \varepsilon k_z}  \\
\Big[ r_{pp} \, \Big|\mathbf{p} \cdot &\hat{\mathbf{E}}_p + \mu \, \mathbf{m} \cdot \hat{\mathbf{H}}_p \, \Big|^2 + \, r_{ss}\, \Big| \, \mathbf{p} \cdot \hat{\mathbf{E}}_s + \mu \, \mathbf{m} \cdot \hat{\mathbf{H}}_s\Big|^2\\
+ \, r_{ps}\, \Big( \, &\mathbf{p} \cdot \hat{\mathbf{E}}_p + \mu \, \mathbf{m} \cdot \hat{\mathbf{H}}_p \, \Big)^*
\Big( \, \mathbf{p} \cdot \hat{\mathbf{E}}_s + \mu \, \mathbf{m} \cdot \hat{\mathbf{H}}_s \, \Big)\\
+ \, r_{sp}\, \Big( \, &\mathbf{p} \cdot \hat{\mathbf{E}}_s + \mu \, \mathbf{m} \cdot \hat{\mathbf{E}}_s \, \Big)^*
\Big( \, \mathbf{p} \cdot \hat{\mathbf{E}}_p + \mu \, \mathbf{m} \cdot \hat{\mathbf{H}}_p \, \Big)\Big]\bigg\}
\end{split}
\end{align*}

\noindent This cross polarization expression may prove useful for readers looking to analytically describe forces above an anisotropic or non-reciprocal planar surface. 


%

\end{document}